\pdfoutput=1
\documentclass[manuscript,screen,nonacm]{acmart}
\acmJournal{TOSEM}
\acmVolume{1}
\acmNumber{1}
\acmArticle{1}
\acmMonth{1}
\acmYear{2026}
\setcopyright{none}

\usepackage[most]{tcolorbox}
\usepackage{xcolor}

\definecolor{RQBlue}{RGB}{24,79,130}
\definecolor{RQLightBlue}{RGB}{242,247,252}

\newtcolorbox{keymsg}[1][]{%
    enhanced,
    breakable,
    colback=RQLightBlue,
    colframe=RQBlue,
    coltitle=white,
    fonttitle=\bfseries,
    fontupper=\itshape,
    boxrule=0.8pt,
    arc=2pt,
    outer arc=2pt,
    left=7pt,
    right=7pt,
    top=5pt,
    bottom=5pt,
    before skip=7pt,
    after skip=7pt,
    attach boxed title to top left={
        xshift=7pt,
        yshift*=-\tcboxedtitleheight/2
    },
    boxed title style={
        colback=RQBlue,
        colframe=RQBlue,
        boxrule=0pt,
        arc=1.5pt,
        left=5pt,
        right=5pt,
        top=2pt,
        bottom=2pt
    },
    #1
}

\usepackage{xcolor}
\usepackage{enumitem}
\usepackage{amsmath,amssymb}

\usepackage{changepage}
\usepackage{framed}

\makeatletter
{\par\unskip\endMakeFramed}
\makeatother
\definecolor{formalshade}{HTML}{F9E3DF}

\definecolor{darkblue}{rgb}{0.2, 0.2, 0.2}

\usepackage{tikz}

\usepackage{tabularx}

\usepackage{tikz}
\usetikzlibrary{arrows.meta, positioning, shapes.geometric, fit}
\usepackage{booktabs}

\usepackage{url}
\usepackage{pgfplots}
\pgfplotsset{compat=1.18}
\usepackage{amsmath,amssymb,amsfonts}
\usepackage{tikz}
\usetikzlibrary{arrows.meta, positioning}
\usepackage{listings}
\usepackage{xcolor}
\makeatletter
\let\old@lstKV@SwitchCases\lstKV@SwitchCases
\def\lstKV@SwitchCases#1#2#3{}
\makeatother
\usepackage{lstlinebgrd}
\makeatletter
\let\lstKV@SwitchCases\old@lstKV@SwitchCases
\lst@Key{linebackgroundcolor}{}{\def\lst@linebgrdcolor{#1}}
\lst@Key{linebackgroundsep}{0pt}{\def\lst@linebgrdsep{#1}}
\lst@Key{linebackgroundwidth}{\linewidth}{\def\lst@linebgrdwidth{#1}}
\lst@Key{linebackgroundheight}{\ht\strutbox}{\def\lst@linebgrdheight{#1}}
\lst@Key{linebackgrounddepth}{\dp\strutbox}{\def\lst@linebgrddepth{#1}}
\makeatother
\usepackage{algorithmic}
\usepackage{graphicx}
\usepackage{booktabs}
\usepackage{textcomp}
\usepackage{tabularx}
\hypersetup{hidelinks} 
\usepackage{xcolor}
\usepackage{subcaption}

\lstdefinelanguage{JavaScript}{
  morekeywords={var, let, const, function, return, if, else, for, while,
                new, typeof, this, null, true, false, in, of, require},
  sensitive=true,
  morecomment=[l]{//},
  morecomment=[s]{/*}{*/},
  morestring=[b]',
  morestring=[b]"
}

\begin{document}

\title[Preserve the Vulnerability, Break the Alert]{Preserve the Vulnerability, Break the Alert: Testing the Robustness of CodeQL Taint Tracking with Natural Rewrites}

\author{Zhezhen Cao}
\email{zcao24@ncsu.edu}
\affiliation{%
  \institution{North Carolina State University}
  \city{Raleigh}
  \state{NC}
  \country{USA}}

\author{Weiyuan Ding}
\email{wding8@ncsu.edu}
\affiliation{%
  \institution{North Carolina State University}
  \city{Raleigh}
  \state{NC}
  \country{USA}}

\author{Bowen Xu}
\email{bxu22@ncsu.edu}
\affiliation{%
  \institution{North Carolina State University}
  \city{Raleigh}
  \state{NC}
  \country{USA}}

\renewcommand{\shortauthors}{Cao et al.}

\begin{abstract}
Static application security testing (SAST) tools such as CodeQL serve as
security gates in continuous integration and code review, and a gate is
also read in reverse. When a previously reported alert disappears after a
change, developers and the tools around them often take the vulnerability
to be gone. That inference holds only if the analyzer is robust to changes
that keep the vulnerability. We test this robustness for CodeQL's
JavaScript taint tracking with natural, vulnerability-preserving rewrites.
A large language model, given only the alert CodeQL exposes and no
knowledge of its implementation, rewrites one function along a reported
flow as a plausible maintenance change, and we verify by manual audit and
controlled experiments that the flaw survives and the alert does not. On
172 flows that CodeQL detects in real npm vulnerabilities, three rewrites
per flow silence 26.2\% of them, at about \$0.51 per silenced flow. The
same generator, asked only to refactor the same functions with no
knowledge of the alert, silences none of them. Every
silencing rewrite traces to one of seven gaps at the library-modeling
boundary, and 64 of the 70 rewrites cross a gap inside an API CodeQL
already models. The
largest is the WHATWG URL constructor, which propagates taint to a URL's
query components but not to its \texttt{pathname}. The boundary is not
specific to CodeQL. 74\% of the silencing rewrites also silence an
independently built taint analyzer, and 53\% escape three analyzers
combined. A disappearing alert after an ordinary-looking change is
therefore weak evidence that a vulnerability has been fixed, the gaps
point SAST developers to narrow and concrete repair targets, and
robustness to natural rewrites belongs alongside detection rate in how
taint analyzers are evaluated.
\end{abstract}



\maketitle

\section{Introduction}
\label{sec:introduction}

Static application security testing (SAST) increasingly serves as a security
gate in modern software development. Tools such as CodeQL~\cite{codeql-tool},
which powers GitHub code scanning and provides a mature suite of analyses
based on interprocedural dataflow and taint tracking~\cite{codeql-ql},
analyze source code during continuous integration and pull-request review,
where a reported vulnerability can block a contribution from being merged.
The gate is also read in reverse. When a previously reported alert
disappears after a change, developers and the tools around them often take
that as a sign that the vulnerability is gone.
Automated alert-fixing assistants are judged successful when the re-scan is
clean and the tests still pass~\cite{gazit2024fixing,copilot-autofix}, and
developers who edit a flagged function tend to stop once the alert
clears~\cite{imtiaz2019coverity}.

That inference holds only if the analyzer is robust to changes that keep the
vulnerability. A taint analyzer tracks data from sources to sinks through
the code it sees and relies on hand-written summaries for the library calls
it does not analyze, which gives it a \emph{modeling boundary}, the set of
constructs through which it propagates taint~\cite{livshits2015soundiness}.
A change that keeps the vulnerable behavior but moves the flow across that
boundary makes the alert disappear with the flaw intact. Such a change needs
no intent. It is a by-product of any workflow whose stopping criterion is a
clean re-scan, whether the editor is an alert-fixing assistant, a developer
trying alternatives until the alert clears, or a contributor who keeps the
vulnerability deliberately. Large language models (LLMs) make such changes
cheap and varied. They produce idiomatic, function-level rewrites without
any knowledge of the analyzer's implementation, so the question is no longer
whether such rewrites exist but how often ordinary ones cross the boundary.

\begin{figure}[t]\centering
\begin{minipage}[t]{0.48\textwidth}\centering
\begin{lstlisting}[language=JavaScript,frame=tb,basicstyle=\ttfamily\footnotesize,
  showstringspaces=false,breaklines=false,columns=fullflexible,keepspaces=true,
  keywordstyle=\color{blue!65!black}\bfseries,
  commentstyle=\color{gray!55!black}\itshape,
  stringstyle=\color{purple!65!black},
  escapeinside={(*}{*)},
  linebackgroundcolor={\ifnum\value{lstnumber}=4\color{red!15}\fi}]
function getPsInfo(param, callback) { // (*\textcolor{red!75!black}{\textbf{source}}*)
  var pid = param.pid;
  var cmd = "ps auxw | grep " + pid + " | ..."
  exec(cmd, function (err, out) { ... }); // (*\textcolor{red!75!black}{\textbf{sink}}*)
}
\end{lstlisting}
{\footnotesize\textbf{Original.} CodeQL tracks
\texttt{param}\,$\to$\,\texttt{pid}\,$\to$\,\texttt{cmd} \\
and reports the flow.}
\end{minipage}\hfill
\begin{minipage}[t]{0.48\textwidth}\centering
\begin{lstlisting}[language=JavaScript,frame=tb,basicstyle=\ttfamily\footnotesize,
  showstringspaces=false,breaklines=false,columns=fullflexible,keepspaces=true,
  keywordstyle=\color{blue!65!black}\bfseries,
  commentstyle=\color{gray!55!black}\itshape,
  stringstyle=\color{purple!65!black},
  escapeinside={(*}{*)},
  linebackgroundcolor={%
    \ifnum\value{lstnumber}=4\color{green!18}\fi
    \ifnum\value{lstnumber}=5\color{green!18}\fi
    \ifnum\value{lstnumber}=6\color{green!18}\fi}]
function getPsInfo(param, callback) { // (*\textcolor{red!75!black}{\textbf{source}}*)
  var pid = param.pid;
  var cmdParts = ['ps auxw', 'grep ' + pid,
                  "grep -v 'grep'"];
  var cmd = cmdParts.join(' | ');
  exec(cmd, function (err, out) { ... }); // (*\textcolor{red!75!black}{\textbf{sink}}*)
}
\end{lstlisting}
{\footnotesize\textbf{Rewrite.} \texttt{cmd} is reconstructed identically
at runtime. \\
CodeQL loses the taint at
\texttt{join} and no longer reports it.}
\end{minipage}
\caption{An LLM rewrite of a real command-injection
vulnerability~\cite{motivating} silences CodeQL's detection.}
\label{fig:taint_flow}
\end{figure}

Figure~\ref{fig:taint_flow} shows one. The original function concatenates a
request parameter into a shell command, and CodeQL reports the flow from
\texttt{param} to \texttt{exec}. The rewrite assembles the same command from
an array with \texttt{join(' | ')}, a change that reads as a readability
refactoring and reconstructs the identical string at runtime. CodeQL's
command-injection query models \texttt{join} only with a single-space
separator, so the taint stops at the \texttt{join} call and the alert
disappears while the vulnerability remains.

This failure mode is not well characterized by prior work. Studies of
semantics-preserving transformations against code analysis use conspicuous
changes to defeat learned detectors or obfuscate code until a SAST tool
reports clean~\cite{moser2007limits,yang2022natural,liu2024eatvul,pagano2026obfuscating},
and studies that test static analyzers drive the transformations with fixed
rules or synthetic
programs~\cite{klinger2019differentially,zhang2023statfier,bonett2018discovering}.
Neither says how often \emph{natural}, vulnerability-preserving rewrites of
real vulnerable code silence a production taint analyzer, where in its
modeling boundary the silenced flows disappear, or whether the same rewrites
silence independently designed analyzers. The distinction matters
operationally. Conspicuous obfuscation is likely to be rejected in review,
whereas a small readability refactoring or a conventional library call may
be merged precisely because it appears benign.

We study this question as a robustness property of the analyzer. For a
program in which CodeQL reports a flow, a rewrite of one function along that
flow that preserves the function's interface and its vulnerable behavior
should leave the flow reported. We instantiate the rewrite with an LLM that
is \emph{alert-aware but implementation-agnostic}. It sees the source, sink,
and flow that CodeQL exposes and may change one function, but cannot inspect
or alter CodeQL's queries, models, or configuration. A rewrite is a
\emph{silencing rewrite} when it preserves the vulnerable source-to-sink
behavior, CodeQL no longer reports the targeted flow, and the change reads
as plausible maintenance rather than an evasion artifact. Deleting the sink,
sanitizing the input, or otherwise fixing the flaw is excluded. We ask:

\begin{keymsg}
How robust is CodeQL's taint tracking to natural, vulnerability-preserving
rewrites, where in its modeling boundary do the silenced flows disappear,
and is that boundary specific to CodeQL?
\end{keymsg}

To answer this question, we develop a three-stage pipeline over
VulcaN~\cite{vulcan}, a dataset of confirmed vulnerabilities in real-world
npm packages annotated at the \texttt{(source, sink)} level, a granularity
few datasets provide. We rerun CodeQL to recover the flows it detects,
generate three independent function-local rewrites per flow under explicit
vulnerability-preservation and naturalness constraints, and rerun CodeQL to
verify each candidate, confirming by manual audit that the vulnerability
survives and attributing each silenced flow to one modeling gap through
controlled experiments that restore one construct at a time. We then apply
the same rewrites, unchanged, to ODGen and njsscan, which represent
independently implemented semantic and pattern-based
analyses~\cite{odgen,njsscan}.

Across 172 vulnerable flows and 514 valid rewrite attempts, spanning five
CWEs but dominated by path traversal and command injection, three attempts
per flow silence CodeQL's targeted flow on 26.2\% of the flows it
originally detects, at about \$0.51 of model inference per silenced flow.
Ordinary language-level transformations, such as introducing intermediate
variables or destructuring values, almost never silence CodeQL. Every
silencing rewrite instead traces to one of seven gaps at the
library-modeling boundary, and 64 of the 70 rewrites cross a gap
\emph{inside an API that CodeQL already models}. The single largest gap is the WHATWG URL constructor, which
carries taint to a URL's query components but not to its
\texttt{pathname}, and it alone accounts for 43 of the 70. The boundary is
not specific to one analysis implementation. Among the silencing rewrites
in ODGen's scope, 74\% also silence ODGen despite its separate
object-dependence design, and across CodeQL, ODGen, and njsscan, 53\% of
the silencing rewrites in the common scope escape all three. A
disappearing SAST alert is therefore weaker evidence than development
workflows assume. It may mean that the program has crossed the analyzer's
modeling boundary rather than that the vulnerability has been removed.

This paper makes the following contributions:

\begin{itemize}[wide=4pt]
    \item \textbf{A robustness formulation and pipeline for taint
    analyzers.} We cast robustness to natural, vulnerability-preserving
    rewrites as a metamorphic relation, instantiate the rewrites with an
    alert-aware but implementation-agnostic LLM, and verify detection loss
    and vulnerability preservation together, with controlled experiments
    that attribute each silenced flow to one modeling gap, rather than
    inferring a fix from analyzer silence.
    \item \textbf{An empirical measurement on real npm vulnerabilities.} On
    172 flows CodeQL detects, three natural rewrites per flow silence
    26.2\% at about \$0.51 per silenced flow, with every case manually
    verified, while the same generator refactoring the same functions
    without the alert silences none.
    \item \textbf{A root-cause and generality analysis.} We trace the 70
    silencing rewrites to seven modeling gaps, 64 of them inside APIs
    CodeQL already models and all of them still present in the current
    release, relate them to a taxonomy of rewrite strategies, and
    show by applying the rewrites unchanged to ODGen and njsscan that the
    boundary is shared across analysis designs rather than specific to
    CodeQL.
\end{itemize}

\section{Background}
\label{sec:background}

\subsection{Taint Tracking and the Modeling Boundary}
\label{subsec:bg-taint}

Static taint analysis determines whether data from an untrusted \emph{source},
such as an HTTP request field, can reach a security-sensitive \emph{sink}, such
as \texttt{fs.readFile} or \texttt{child\_process.exec}, without passing through
an effective sanitizer~\cite{taj}. The analysis has two parts. Within the code
the analyzer sees, it follows assignments, parameters, return values, and
property reads and writes, and for taint it additionally follows operations
that derive a new value from a tainted one, such as string concatenation or
substring extraction. Local flow stays within one function. Global flow
crosses calls and returns, which security queries require, at the cost of
lower precision~\cite{codeql-docs-dataflow}. Sanitizers are modeled as
\emph{barriers}: nodes at which tracking stops, or guards that stop it on the
branch where a validity check succeeds.

The second part concerns calls that leave the analyzed code: functions of the
language runtime, the standard library, and third-party packages, whose bodies
the analysis does not traverse, either because they are unavailable or because
analyzing them would not scale. For these calls it relies on
\emph{summaries}, hand-written models that
state how a call propagates taint among its arguments, receiver, and result. A
call without a summary propagates nothing. This design is a deliberate trade of
soundness for precision and scalability, and it gives every taint analyzer a
\emph{modeling boundary}: the set of constructs and API behaviors through
which it propagates taint. When a flow crosses that boundary, the analyzer
drops it, although the tainted value still reaches the sink when the program
runs.

An API the analyzer never models is therefore a known blind spot: it is the
premise of the soundiness position~\cite{livshits2015soundiness} and the
motivation for inferring missing
specifications~\cite{merlin,seldon,inspectjs,iris2025}. The boundary,
however, does not simply separate modeled APIs from unmodeled ones. A summary
may describe some outputs of an operation but not others, so the analyzer can
lose a flow inside an operation it otherwise models. Which of the two kinds
of gap ordinary code changes tend to reach is an empirical question, and one
this paper takes up in Section~\ref{subsec:rq2}.

\subsection{CodeQL's JavaScript Taint Analysis}
\label{subsec:bg-codeql}

CodeQL~\cite{codeql-ql,codeql-tool} compiles a program into a relational
database and expresses analyses as declarative queries over it. Security
queries are organized into suites. Each query targets one vulnerability class,
for example \texttt{js/path-injection} for CWE-22, and declares its sources,
sinks, barriers, and additional taint steps on top of a shared global
data-flow engine~\cite{codeql-docs-dataflow}. The engine handles ordinary
language constructs, including interprocedural flow through calls, returns,
closures, and property accesses, so taint rarely escapes it by passing through
plain code. Library behavior is where modeling is explicit. CodeQL's
JavaScript \emph{library models} summarize how built-in and npm APIs propagate
taint, written either as QL classes or as extensible models-as-data
tables~\cite{codeql-mad}. Models are pattern-based: a summary applies to a
particular callee, call shape, and argument position, and a call outside that
shape receives no summary. Some models are also specific to a query. For
instance, the path-injection query treats \texttt{Array.prototype.join} as
string concatenation only when the separator is the empty string, whereas the
command-injection query recognizes it with a single-space separator
(Section~\ref{subsec:rq2}). Together, these shapes define CodeQL's modeling
boundary.

A result of a path query is a \texttt{(source, sink)} pair accompanied by a bounded
number of representative \emph{code flows}, four by default, serialized in
SARIF~\cite{sarif}. The code flow is the unit of analysis in our study. CodeQL
thus reports that a sink is reachable from a source, not every route by which
it is reachable. We say that CodeQL \emph{detects} an annotated vulnerability
when it reports a flow matching the annotated source and sink, with the
matching made precise in Section~\ref{subsec:match}.

\subsection{Robustness as a Testing Property}
\label{subsec:bg-robustness}

A taint analyzer has no complete oracle: for an arbitrary program, no
reference says which alerts it should report. Metamorphic testing sidesteps a
missing oracle by checking a relation between the outputs on related
inputs~\cite{segura2016survey}, and it has been applied to program analyzers
by comparing analyzers against one another or against themselves on
transformed programs~\cite{cuoq2012testing,klinger2019differentially,taneja2020testing}.
The property we test is one such relation. Let $A$ be a fixed analyzer
configuration, $P$ a program in which $A$ reports a flow $f$, and $T$ a
transformation of one function along $f$ that preserves the function's
interface and the vulnerable flow. Robustness requires that $A$ still report
$f$ on $T(P)$. A violation, in which $f$ is preserved but no longer reported,
is a false negative of $A$ on $T(P)$: a vulnerability that the transformed
program contains and the analyzer misses. The transformation does not create
this miss but exposes one that $A$ would equally make on a program written
that way from the start. The relation is informative only if $T$ is
restricted. An arbitrary transformation, such as obfuscation, violates it
trivially and says little about practice. We therefore restrict $T$ to changes
that could plausibly appear in routine maintenance and call a violating change
of this kind a \emph{silencing rewrite}. Section~\ref{sec:threat} defines
both terms. We instantiate $T$ with an LLM, measure how often the relation is
violated on real npm vulnerabilities, and trace each violation to the part of
the modeling boundary it crosses.

\section{Problem Setting}
\label{sec:threat}

\textbf{Setting.}
We consider an open-source npm package that runs CodeQL in continuous
integration and reviews contributed changes before they merge, typically under
limited reviewing resources~\cite{wermke2023always}. CodeQL reports a genuine
source-to-sink flow in the current code. A later change makes the re-scan
clean, and the alert disappears. Two kinds of change produce this outcome. The
first removes the vulnerability. The second keeps the vulnerable behavior but
moves the flow outside what CodeQL models, so the tool stops reporting a flaw
that is still present. We study the second kind: how often ordinary-looking
changes produce it, and why.

\textbf{Natural, vulnerability-preserving rewrites.}
The object of study is a rewrite of a single function along a reported flow
that (i) preserves the function's externally visible interface, (ii) preserves
the vulnerable behavior, and (iii) reads as routine maintenance, such as a
readability refactoring or a conventional API substitution, rather than as an
evasion artifact. We call such a change a \emph{natural rewrite}. Changes that
delete the sink, sanitize the input, or otherwise remove the flaw are fixes,
not rewrites in our sense, and are excluded by the verification procedure of
Section~\ref{subsec:verify}.

\textbf{Who produces the rewrite.}
Natural rewrites require no intent to deceive. They are a by-product of any
workflow whose completion criterion is a clean re-scan rather than a verified
fix. In such a workflow, a change that keeps the vulnerable behavior but moves
the flow outside CodeQL's models is indistinguishable from a fix, because both
make the alert disappear. We distinguish three producers of such changes.
The first is an automated alert-fixing assistant. GitHub's Copilot Autofix
proposes a patch for a reported CodeQL alert, and GitHub evaluates a suggestion
by merging it and re-running code scanning together with the repository's
tests, counting the suggestion as successful when the alert is gone, no new
alert appears, and the tests still pass~\cite{gazit2024fixing,copilot-autofix}.
None of these checks establishes that the vulnerability is gone, since the
tests exercise functionality rather than the flaw. Repair systems that start
from a static-analysis alert, in research and in industrial tools alike, are
typically judged the same way, by whether the analyzer still reports the
flaw~\cite{pearce2023examining,berabi2024deepcode}.
The second producer is a developer working by hand. Analyzers stop reporting
alerts for many reasons other than a fix, and in practice such alerts are
nonetheless recorded as fixed~\cite{imtiaz2019coverity}, while suppressing an
alert is itself a common way of making it disappear~\cite{hu2025suppressed}. A
developer who addresses an alert by editing the flagged function and
re-scanning may therefore stop once the alert clears, especially when the
alert is hard to interpret~\cite{johnson2013why}.
The third producer is a contributor who deliberately keeps the vulnerability
while passing the CodeQL gate. We treat this case as a corollary of the first
two rather than as the motivation. It is the only one that requires an
assumption about intent, and the measurement already covers it, because a
deliberate contributor sees exactly the same information as the other two.
Section~\ref{subsec:implications} returns to this case.
All three producers see what CodeQL exposes, namely the alert with its source,
sink, and flow, and can re-run CodeQL on candidate changes, but none inspects
or modifies CodeQL's queries, library models, or configuration. We therefore
model the rewriter as \emph{alert-aware but implementation-agnostic}.

\textbf{Rewrite generator.}
We instantiate the rewriter with a single LLM (model and sampling in
Section~\ref{subsec:impl}). It receives the reported flow and the enclosing
functions, is instructed to preserve the vulnerable behavior and to produce a
maintenance-style change, and is never given CodeQL's implementation or model
definitions. The rates we report therefore characterize this bounded
generator, alert-aware, implementation-agnostic, and limited to a fixed
attempt budget. They are a floor rather than a ceiling: a generator that could
read the library models would target their gaps directly, and more expertise
or more attempts could raise the rates further.

\textbf{Silencing rewrites.}
A natural rewrite is a \emph{silencing rewrite} when all three conditions hold
after the change: (1)~\emph{vulnerability preservation}: data from the original
source still reaches the original sink without an effective sanitizer;
(2)~\emph{detection loss}: CodeQL no longer reports the targeted source-to-sink
flow under the flow-matching procedure of Section~\ref{sec:approach}; and
(3)~\emph{naturalness}: the rewrite meets the maintenance-change criteria
above, as operationalized in Section~\ref{subsec:audit}. Conditions (1)
and~(2) together are the violation of the relation in
Section~\ref{subsec:bg-robustness}, with $T$ the rewrite and $f$ the targeted
flow. Detection loss is defined at the flow level: CodeQL may still report a
different flow in the same function or file. We also report the stricter vulnerability-level criterion,
under which no finding of the targeted rule remains in the rewritten function.

\textbf{What we hold fixed.}
Each rewrite changes exactly one function, keeps its externally visible
interface, must be syntactically valid, and must not use overt obfuscation such
as assembling identifiers from string fragments. CodeQL's version, query suite,
and configuration are identical before and after the rewrite, so any lost
detection is attributable to the code change alone.

\section{Approach}
\label{sec:approach}

\begin{figure}[t]\centering
\includegraphics[width=\textwidth]{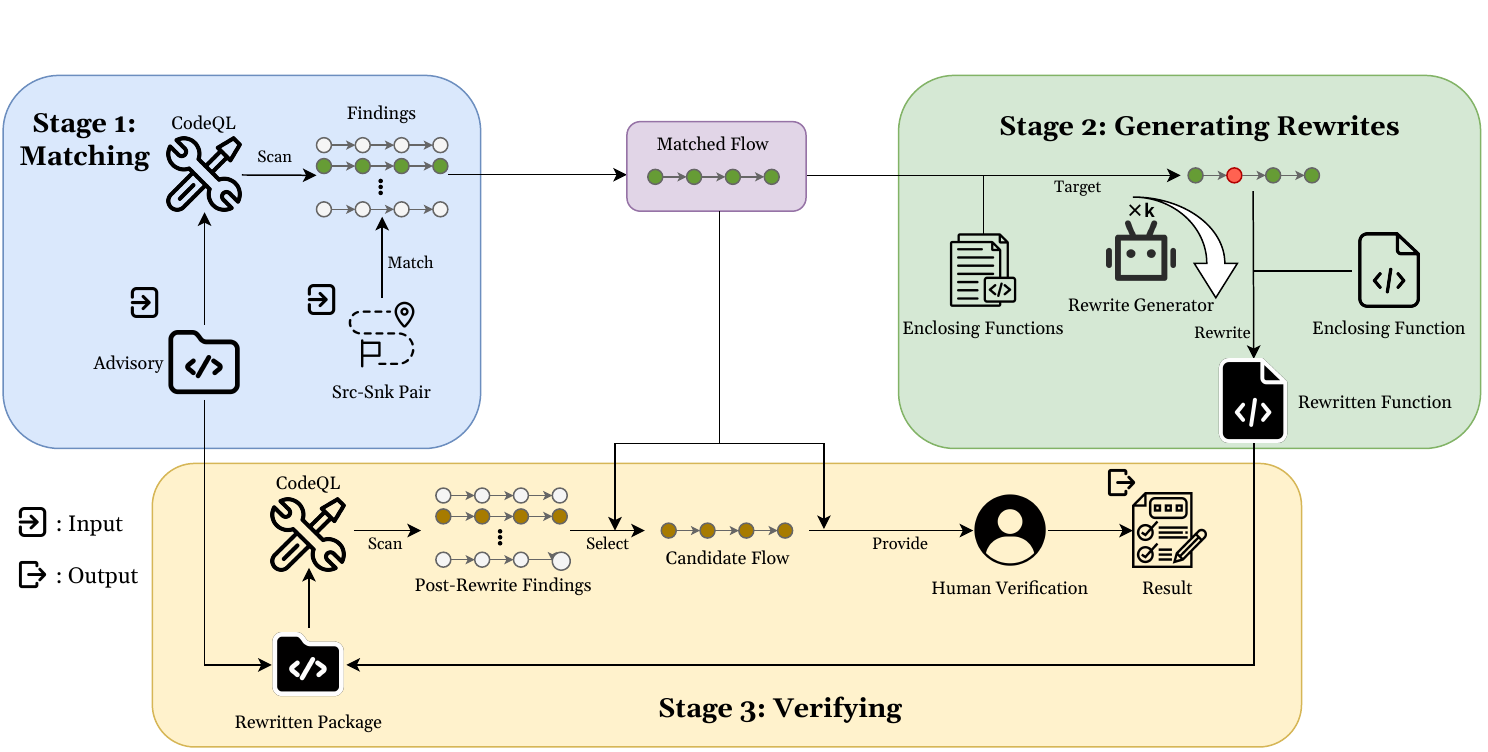}
\caption{Overview of the three-stage pipeline: matching annotated vulnerabilities to the flows CodeQL reports (Stage~1), generating $k$ function-level rewrites per matched flow (Stage~2), and re-running CodeQL and verifying each rewrite (Stage~3).}
\label{fig:pipeline}
\end{figure}

Figure~\ref{fig:pipeline} summarizes the three-stage pipeline we present, which converts
human-annotated advisories into manually verified rewrite outcomes. Stage~1 builds a CodeQL database for each advisory, runs
the CWE-specific queries, and aligns CodeQL's reported code flows with the
annotated \texttt{(source, sink)} pair. Only matched flows proceed, because a
flow CodeQL does not report in the original program cannot be silenced.
Stage~2 instantiates the rewrite generator of
Section~\ref{sec:threat} with an LLM and generates $k$ function-level rewrites
for each matched flow under explicit vulnerability-preservation and
naturalness constraints. Stage~3 re-runs
CodeQL, aligns the post-rewrite findings with the original target, and presents evidence for human verification.

Figure~\ref{fig:taint_flow} shows the outcome Stage~3 confirms for one
command-injection flow: the targeted \texttt{param}-to-\texttt{exec} flow
disappears after the array-assembly rewrite, while the tainted value still
reaches \texttt{exec}.

\subsection{Stage 1: Matching Detected Flows}
\label{subsec:match}
Stage~1 identifies target flows by intersecting confirmed vulnerabilities
with the code flows that CodeQL reports in our environment. A CodeQL alert
alone is insufficient because static analyzers may report false positives, so
we use the dataset's human-annotated \texttt{(source, sink)} pairs as the
reference for genuine vulnerabilities. Conversely, an annotated vulnerability
that CodeQL does not detect provides no alert to silence. We therefore
retain only reported code flows that align with an annotation; pairs that
fail this alignment appear in the \emph{Unmatched} column of
Table~\ref{tab:dataset}.

For each advisory, we build a CodeQL database, run the CWE-specific queries
(Section~\ref{sec:setup}), and parse the first code flow CodeQL reports for
each result, its representative path, as an ordered sequence. Because the annotation specifies only two endpoints and an advisory
may contain unrelated findings in the same file, the matching procedure must
determine which reported flow, if any, corresponds to the annotated
vulnerability. Under the default rule, the first step of the flow must fall
within a fixed line tolerance of the annotated source, and one of the flow's
final steps must lie on the exact annotated sink line: the source often refers
to the same entry point with a small line offset, whereas the dangerous sink
operation should coincide exactly. If multiple flows satisfy the rule, we choose the one whose matched sink is closest to the end of the reported
path, then the shortest.

Command injection (CWE-77/78) and XSS (CWE-79) use a relaxed rule for two
recurring patterns. First, the annotated source may correspond to any flow step
before the sink rather than only the first, since the annotation may identify a
library entry-point function while CodeQL begins propagation at a deeper
operation. Second, the annotated sink may fall within the same line tolerance,
since CodeQL can report the command or markup construction a few lines before
the dangerous operation. The matched source must still precede the matched sink,
preserving the direction of the reported flow.

\subsection{Stage 2: Generating Vulnerability-Preserving Rewrites}
\label{subsec:rewrite}

Stage~2 uses an LLM to search a broad space of idiomatic, function-level
transformations while enforcing the preservation and naturalness
requirements of Section~\ref{sec:threat}. A fixed set of mutation rules would restrict the
search to predetermined surface forms; the LLM can instead propose varied
refactorings and API substitutions. We separate generation into two turns: the
first selects a location along the matched flow, and the second rewrites the
corresponding enclosing function.

\textbf{Turn~1: selecting a target location.} The model receives the CWE and
the matched CodeQL flow. Because CodeQL may revisit one line through several
subexpressions, we collapse the reported steps into distinct locations (unique
\texttt{file:line} pairs) and provide the enclosing function for each. The
model selects one location whose surrounding computation it will rewrite and
briefly explains how an equivalent formulation may alter the dataflow visible
to CodeQL. It also produces a short strategy label and a one-line \emph{commit
message} that states a plausible maintenance rationale, such as improving
readability.

\textbf{Turn~2: producing the rewrite.} The model receives the selected
location's enclosing function and the commit message and returns a replacement
for that function. It additionally emits a one-sentence source-to-sink trace and
a short label describing the applied transformation. The trace is shown to the
auditor as supporting context rather than accepted as proof of vulnerability
preservation, while the strategy label contributes to the taxonomy analyzed in
Section~\ref{subsec:rq2}. After applying the replacement, we run
\texttt{node -{}-check}. If parsing fails, we return the error to the model and
allow up to two repair calls. Attempts that remain syntactically invalid are
marked invalid; Section~\ref{subsec:metrics} specifies how invalid attempts are
treated by pass@$k$ and SR.

\textbf{Generation constraints.} The prompt encodes both requirements of
Section~\ref{sec:threat}: the rewrite must preserve the original
source, sink, and vulnerable behavior, and must read as a plausible maintenance
change rather than an artifact aimed at the analyzer. Stage~3 marks an attempt invalid
if the applied rewrite violates any of these constraints.

We repeat the two-turn procedure $k$ times per flow. Each attempt
starts from the same original function and matched flow, and earlier rewrites or
their post-rewrite CodeQL outcomes are not supplied to later attempts, so
pass@$k$ records whether at least one of these independent attempts silences
the alert. The benign baseline of RQ1 is a variant of this stage that skips
Turn~1 and gives Turn~2 neither the flow nor any instruction about the
vulnerability or the alert (Section~\ref{subsec:impl}).

\subsection{Stage 3: Verifying Silencing}
\label{subsec:verify}

Stage~3 combines automated flow alignment with manual verification: automation
narrows the evidence presented to the auditors but does not decide whether a
rewrite is a silencing rewrite. This distinction is necessary because a missing target may
result from broken code or a changed vulnerability, while a surviving CodeQL
finding may correspond to a different flow in the same function.

\textbf{CodeQL re-analysis and candidate alignment.} We re-run CodeQL on each
rewritten package using the same queries and analysis procedure as in Stage~1.
A rewrite may shift line numbers and may cause CodeQL to report several
findings, so we automatically compare every post-rewrite finding, including
each of its reported code flows, with the original target.
The comparison compensates for line shifts, aligns the two step sequences using
the Needleman--Wunsch algorithm, checks endpoint correspondence, and surfaces
the best candidate flow, or reports that no candidate was found. The alignment
also records whether the rewrite changed the source, the sink, or an
intermediate location for the analysis in Section~\ref{subsec:rq2}. Its output
is a prioritization aid; the final classification remains a human judgment.

\textbf{Audit interface.} A web application presents, for each attempt, the
original matched flow, the rewritten function and diff, the best post-rewrite
candidate, and the model's self-reported trace in a single view, so auditors
evaluate each attempt under consistent evidence.

\textbf{Human verification.} The auditor first determines whether CodeQL still
reports the targeted flow, then verifies vulnerability preservation:
when the target survives, the reported source-to-sink flow provides direct
evidence; when the target disappears, the auditor manually traces whether data
from the original source can still reach the original sink without effective
sanitization. The auditor also checks syntactic and semantic validity, the
Stage~2 constraints, and the naturalness criteria from Section~\ref{subsec:audit}.
An attempt is labeled a silencing rewrite only if it meets the three
conditions of Section~\ref{sec:threat}: the targeted flow is no longer
reported, the original vulnerability remains, and the rewrite satisfies the
stated constraints.

Potentially broken rewrites receive an explicit validation control rather than being counted from CodeQL silence alone. In particular, a function-only rewrite may reference a module that the surrounding file does not import. While Section~\ref{sec:threat} restricts a rewrite to one function, we add the missing import at the file scope as a diagnostic control. This ensures we are measuring genuine analysis failures rather than simply broken code. We then re-run CodeQL, retaining the attempt as a silencing rewrite only if
the completed program remains vulnerable and the targeted flow is still
absent; otherwise, the attempt is invalid.

Every confirmed silencing rewrite undergoes controlled experiments to
verify that the disappearance is stable and to identify the analysis
boundary responsible for it. They include restoring the original operation
or varying one part of the rewrite while holding the rest constant,
followed by a fresh CodeQL run, and they end by assigning the case to one
modeling gap, the unit of the root-cause analysis in
Section~\ref{subsec:rq2}. A case is assigned to a gap only when restoring
the blamed construct alone, with the rest of the rewrite kept, brings the
targeted flow back. Because every positive case receives this
additional analysis whereas only a sample of negative cases receives a
second audit (Section~\ref{subsec:audit}), the protocol is an asymmetric
verification cascade rather than symmetric dual coding. It is designed to
minimize spurious silencing labels. It may miss some silencing rewrites
among the negatives, so the measured silencing rate is conservative.

\section{Experimental Setup}
\label{sec:setup}

\subsection{Research Questions}
\label{subsec:rqs}

Our evaluation addresses three research questions:

\begin{itemize}[wide=0pt]
    \item \textbf{RQ1 (Robustness): How often do natural,
    vulnerability-preserving rewrites silence a confirmed CodeQL alert?}
    We measure the fraction of detected flows that a bounded rewrite
    generator silences within $k$ attempts while the vulnerability remains,
    how this fraction varies across vulnerability classes, what each
    silenced flow costs to produce, and how it compares with a benign
    baseline in which the same generator refactors the same functions with
    no instruction about the vulnerability or the alert.
    \item \textbf{RQ2 (Mechanisms): Where in CodeQL's modeling boundary do
    the silenced flows disappear, and which rewrites reach those gaps?} We
    trace every confirmed silencing rewrite to the dataflow or
    library-modeling gap responsible for it, classify the transformation
    strategies the generator applies, and relate the two. We also re-run the
    rewrites under the current CodeQL release and report, per gap, whether
    it persists.
    \item \textbf{RQ3 (Generality): Is the exposed boundary specific to
    CodeQL?} We apply the rewrites unchanged to an independently built
    flow-sensitive taint analyzer and ask whether they silence its alerts
    too, with a pattern-matching scanner that does not track flows as a
    negative control.
\end{itemize}

\subsection{Dataset}
\label{subsec:dataset}

\begin{table}[t]
\centering\small
\caption{Dataset corpus for the five studied CWEs. \emph{Pairs} are VulcaN's
annotated \texttt{(source, sink)} pairs. \emph{No finding} and
\emph{Unmatched} are pairs CodeQL does not report in our setup
(Section~\ref{subsec:match}). \emph{Targets} are the flows rewritten in
Stage~2.}
\label{tab:dataset}
\begin{tabular}{lrrrrr}
\toprule
CWE & Advisories & Pairs & No finding & Unmatched & Targets \\
\midrule
CWE-22 (path traversal)    & 103 & 124 & 4  & 28 & 92 \\
CWE-23 (path traversal)    & 15  & 16  & 2  & 0  & 14 \\
CWE-77 (command inj.)      & 12  & 21  & 0  & 10 & 11 \\
CWE-78 (command inj.)      & 46  & 72  & 1  & 22 & 49 \\
CWE-79 (XSS)               & 18  & 26  & 7  & 12 & 7  \\
\midrule
Total                      & 194 & 259 & 14 & 72 & 173 \\
\bottomrule
\end{tabular}
\end{table}

We draw our subjects from VulcaN~\cite{vulcan}, a dataset of real-world npm
vulnerabilities in which each advisory carries a human-annotated
\texttt{(source, sink)} pair. Our pipeline requires ground truth at this
granularity, since Stage~1 aligns CodeQL flows to annotated endpoints
(Section~\ref{subsec:match}), and VulcaN is one of the few benchmarks that
labels endpoints directly rather than a vulnerable function or file.

We keep the advisories that VulcaN marks at CodeQL's highest-precision tier
(\emph{Score-A}: the correct rule firing on the annotated source and sink)
and target the five CWEs with enough advisories and close
annotation--CodeQL agreement to support per-CWE analysis: path traversal
(CWE-22 and CWE-23), command injection (CWE-77 and CWE-78), and cross-site
scripting (CWE-79). These five CWEs contribute 194 advisories with 259
annotated pairs. The selection was fixed before Stage~2 and is therefore
uninfluenced by rewrite outcomes.

VulcaN's annotations are line-level \texttt{(file, line)} pairs, located
manually and cross-checked by two of its authors, and its Score-A tier was
assigned under CodeQL~2.2.6, whereas we run CodeQL~2.22.4
(Section~\ref{subsec:impl}), which reports flows as expression-level step
sequences. Every pair is therefore revalidated in our environment.
Re-running CodeQL and applying the Stage-1 matching confirms 173 target
flows from 145 of the 194 advisories (Table~\ref{tab:dataset}). The
remaining 86 pairs are not usable as ground truth: for 14, CodeQL reports
no finding of the relevant rules at all, and for 72 it reports findings
whose endpoints do not align with the annotated pair. Both losses fall
hardest on XSS, which starts from the fewest pairs and retains only 7 of
26, so the XSS portion of the corpus is small.
We exclude one flow whose source is minified, since our setting targets
human-readable functions submitted for maintainer review
(Section~\ref{sec:threat}) and minified code falls outside that scope.
Every remaining flow retains at least one valid attempt under the criteria
of Section~\ref{subsec:impl}, giving a final corpus of 172 flows. The
benign baseline of RQ1 uses the same 172 flows. The corpus is dominated by
path traversal (CWE-22 alone yields 92 targets), whereas command injection
via CWE-77 (11) and XSS (7) are thin, so we treat their per-CWE rates as
indicative only (Section~\ref{subsec:ttv}).

\subsection{Metrics}
\label{subsec:metrics}
\textbf{Silencing metrics.}
We report two complementary metrics over the attempts of each flow.
\emph{pass@$k$} is the expected fraction of flows for which at least one of
$k$ independent rewrites is a confirmed silencing rewrite, computed with the
unbiased estimator~\cite{Chen2021EvaluatingLL} over the $n{=}3$ attempts
per flow, where an invalid attempt counts as a failed sample. In the setting
of Section~\ref{sec:threat} the $k$ rewrites are independent edits of the
same flagged function, each judged only by whether the re-scan is clean, so
pass@$k$ is the probability that an alert-driven workflow with a budget of
$k$ edits ends with the alert gone and the vulnerability intact. A reviewer
sees only the change at hand, never a rejected predecessor. The
\emph{Silencing Rate} (SR) is the fraction of valid attempts that are
silencing rewrites, that is, that violate the relation of
Section~\ref{subsec:bg-robustness}. It measures per-attempt effectiveness
and underlies our strategy and control comparisons, as well as the cost we
report as the generator's API spend per flow and per silenced flow. An
attempt is invalid when it breaks a Stage~2 constraint, remains unparseable
after repair, or fails to produce a usable analyzer run. Invalid attempts
are excluded from SR and cannot contribute a success to pass@$k$.

Both metrics use two granularities. At the \emph{flow level} CodeQL no
longer reports the targeted flow, and at the stricter
\emph{vulnerability level} it reports no flow of the targeted rule through
the rewritten function. We use flow-level pass@$k$ as the primary
robustness measure and report SR at both levels. Because several per-CWE
and cross-tool subsets are small, we give 95\% Wilson confidence
intervals for the per-class pass@3 and for the transfer rates.

\textbf{Control metrics.}
The benign baseline of RQ1 is scored with the same two metrics and the same
verification, so its flow-level pass@$k$ is directly comparable with the
main result; a benign rewrite that removes the vulnerability is a fix and
is excluded in the same way.

\subsection{Primary and Transfer Analyzers}
\label{subsec:transfer-targets}
CodeQL is our primary subject for three reasons. As the analyzer behind
GitHub code scanning, it is the one most often placed as a gate in the
workflow of Section~\ref{sec:threat}. Its JavaScript library models are
open and documented, so each silenced flow can be attributed to a specific
modeling gap in RQ2. And it is the strongest flow-based detector on VulcaN,
so the flows it confirms are a demanding starting point.

The two other analyzers serve RQ3, which asks whether the boundary those
gaps expose is specific to CodeQL. The question calls for a transfer study
rather than a head-to-head comparison. Every rewrite is generated against
CodeQL's alert and applied unchanged to the other analyzers, which see
neither the alert nor the rewrite objective. We choose two analyzers that
handle Node.js packages and appear in VulcaN but realize designs different
from CodeQL and from each other. ODGen~\cite{odgen} builds a flow- and
context-sensitive Object Dependence Graph and evaluates
vulnerability-specific graph queries over it, an independently implemented
semantic taint analysis. Whether the same rewrites silence ODGen
separates a boundary in how taint is modeled from one in CodeQL's
implementation. Transfer is measured only on flows ODGen detects in the
original code, so its rate is conditional on that scope. njsscan~\cite{njsscan}, the scanner behind
NodeJsScan, is built on Semgrep and matches syntax-sensitive patterns
around dangerous APIs without reconstructing source-to-sink flows. It
serves as a negative control. Because it does not track flows, comparing
its outcome on rewrites that silence CodeQL with its outcome on rewrites
that do not tells whether the losses are specific to flow reasoning. We run ODGen at revision \texttt{c5157ee} and njsscan
\texttt{v0.4.3} with their default configurations pinned in our artifact.

\subsection{Implementation Details}
\label{subsec:impl}
\textbf{CodeQL configuration.}
We run CodeQL CLI v2.22.4 with the unmodified JavaScript queries from the
\texttt{security-extended} suite, selecting for each advisory the queries
mapped to its CWE. We analyze every original and rewritten package with the
same CodeQL version, query suite, and database-construction procedure, and
apply the flow-matching procedure of Section~\ref{subsec:match} without
tuning it to rewrite outcomes. Because this release postdates the one under
which VulcaN assigned its Score-A tier (Section~\ref{subsec:dataset}),
every target is revalidated by Stage~1 in our environment.
For the persistence analysis of RQ2 we re-run Stage~1, Stage~3, and the
construct-level probes of Section~\ref{subsec:rq2} under CodeQL CLI
v2.27.3, the current release at the time of writing, with the queries at
the matching tag, keeping the rewrites, the query selection, and the
matching procedure fixed. Every comparison is made within the newer
release: the targeted flow is taken from its own Stage~1 output on the
original package, and a rewrite counts as still silencing only if that
flow is absent from its output on the rewritten package, judged over every
reported code flow with the alignment criterion of
Section~\ref{subsec:verify}. The per-case restore controls are not
repeated, since no case changes outcome.

\textbf{Rewrite generator and sampling.}
We instantiate the rewrite generator of Section~\ref{sec:threat} with
Claude Opus~4.7~\cite{anthropic_opus47}, model identifier
\texttt{claude-opus-4-7}, using
one fixed model for the main study because our objective is to
characterize analyzer robustness under a capable generator, not to compare
language models. Each flow receives $k=3$ independent attempts, each
starting in a fresh context with the same original functions, CodeQL flow,
and prompt, so no earlier rewrite influences a later one. Because the
model does not expose non-default sampling parameters through its API, we
use the service's default configuration. The prompts state the objective
of Section~\ref{sec:threat} directly: the rewrite should keep the reported
flow and its vulnerable behavior, read as a maintenance-style change, and
leave CodeQL no longer reporting the flow. They give the model only what
the alert exposes and nothing about CodeQL's queries or library models.
The model performed the task as stated, with no jailbreak or other
circumvention of its safety behavior, and every subject is a publicly
disclosed vulnerability in an open-source package.

\textbf{Benign baseline.}
The benign baseline of RQ1 uses the same model, the same $k$, the same
sampling configuration, and the same function as the main run: the function
the generator rewrote for that flow. It is a single-turn conversation that
skips the location-selection turn, and its prompt contains only the
function and its file path, with no mention of the vulnerability, the
alert, or any analyzer. The instruction is to refactor for readability
and maintainability while preserving behavior and the external
interface. Syntax
checking, replacement, CodeQL analysis, and flow matching are identical
to the main run. On the 172 flows this gives 516 attempts. For the four
flows whose main-run attempts rewrote two different functions, we also
rewrote the second function, which adds 12 attempts that we report
separately. Because no benign rewrite silences the alert, verification
reduces to confirming that CodeQL still reports the original flow. Of
the 528 attempts, an automatic alignment of the reported flow's endpoints
with the original ones settles 436, and the first author inspected the
remaining 92 with the original and rewritten code and flows side by side
and confirmed the alert on every one. We did not score behavior preservation
for the benign set, since it does not affect whether the alert is
reported.

\textbf{Validity and yield.}
After each rewrite we check syntax with \texttt{node -{}-check} under
Node~v20, treating ES-module files as modules, and allow up to two repair
calls (Section~\ref{subsec:rewrite}). An attempt is invalid if it remains
unparseable, violates a Stage~2 constraint, or cannot be analyzed. The 173
matched flows give 519 planned attempts. Removing the minified flow (three)
and excluding two attempts whose rewrite call fails leaves 514 valid
attempts over 172 flows.

\subsection{Audit Protocol}
\label{subsec:audit}
One author audits all 514 valid attempts under the procedure of
Section~\ref{subsec:verify}. A second author independently re-checks every
confirmed silencing rewrite and a random 10\% sample of the remaining
attempts, and disagreements are resolved by joint inspection to obtain the
final labels, a standard independent-checking design for reducing
subjective labeling error~\cite{wohlin2012experimentation,o2020intercoder}.
Because this design exhaustively rechecks positives but only samples
negatives, we do not report a chance-corrected agreement coefficient such
as Cohen's $\kappa$.

The audit also enforces the naturalness criteria of
Section~\ref{sec:threat}: a rewrite is natural when it uses conventional
programming constructs rather than analyzer-specific artifacts and admits a
plausible maintenance rationale. Every attempt in the main run met them.
The auditors apply these criteria knowing each attempt's outcome, so they
establish that the generator produced the kind of change it was instructed
to produce, not that a reviewer would accept it; Section~\ref{subsec:ttv}
returns to this limitation.

For RQ2, the authors derive the transformation taxonomy by open coding the
applied diff and the model's strategy description, iteratively refining the
codebook, assigning each attempt one dominant strategy, and reconciling
ambiguous cases, following established thematic-analysis practice in
empirical software engineering~\cite{cruzes2011recommended}.

\section{Results}
\label{sec:results}

We analyze 514 valid attempts across 172 vulnerable flows. As described in
Section~\ref{subsec:verify}, each attempt is manually audited and each
candidate silencing rewrite is additionally verified by controlled
experiments, with this audit rather than CodeQL's output deciding every
judgment. All retained attempts preserve the vulnerability and meet the
naturalness criteria of Section~\ref{subsec:audit}, so all rates below are
computed over valid, vulnerability-preserving rewrites.

\subsection{RQ1: Robustness of CodeQL to Natural Rewrites}
\label{subsec:rq1}

With three natural rewrites per flow, the generator silences the targeted
CodeQL flow for $26.2\%$ of the evaluated flows (pass@3). Stage~1 keeps
only flows CodeQL reports on the original code, so each silenced flow is a
false negative that CodeQL would equally produce on a program written that
way from the start, not a vulnerability it never caught.

We find that \textbf{silencing increases with the attempt budget} (columns
pass@$k$ for $k\in\{1,2,3\}$ in Table~\ref{tab:bsr}). Overall pass@$k$
rises from $13.6\%$ to $21.1\%$ and $26.2\%$ at $k{=}1, 2, 3$. The second
attempt adds $7.6$ percentage points and the third adds a further $5.0$.
The marginal gain thus decreases, but the curve has not flattened within
our three-attempt budget. Under the workflow of Section~\ref{sec:threat},
pass@$k$ is the probability that alert-driven fixing with a budget of
$k$ edits ends with the alert gone and the flaw intact, so the measured
pass@3 is a lower bound for larger budgets rather than a ceiling.

Moreover, we observe that \textbf{the robustness of CodeQL varies across
vulnerability classes} (the flow-level \textsc{SR} column in
Table~\ref{tab:bsr}). The overall flow-level SR is $13.6\%$, roughly one
in seven valid rewrites. Aggregated by vulnerability family, path traversal
(CWE-22 and CWE-23) is more exposed than command injection (CWE-77 and
CWE-78), with attempt-level SRs of $16.9\%$ ($53/314$) and $8.9\%$
($16/179$). The gap also holds between the two dominant single classes,
CWE-22 at $17.3\%$ and CWE-78 at $7.5\%$. We trace this gap to the
specific modeling weaknesses each class can reach in
Section~\ref{subsec:rq2}.

The same pattern holds across the attempt budgets. The ordering is stable.
At every $k$ CWE-22 sits above the overall rate and CWE-78 below it, and
their gap widens from $9.7$ percentage points at $k{=}1$ to $18.7$ at
$k{=}3$ (Table~\ref{tab:bsr}). Because the two classes account for $81\%$
of the evaluated flows, they largely determine the aggregate trend, whereas
CWE-77 and CWE-79, with only 11 and 7 flows, are indicative rather than
reliable class-level comparisons. Class, not flow length, is what tracks
silencing. The attempt-level SR stays flat up to 3--5 distinct
\texttt{file:line} locations and halves only for the longest paths
($\geq 6$), which are disproportionately the more robust command-injection
and XSS flows.

\begin{table}[t]
\centering
\caption{Per-class silencing results. $pass@k$ is flow-level, \emph{Att.}
counts valid rewrite attempts, \emph{n} counts confirmed silencing
rewrites, and \emph{CI} is the 95\% Wilson interval of pass@3. The last
row is the benign baseline of Section~\ref{subsec:impl}.}
\label{tab:bsr}
\small
\begin{tabular}{@{}lrrrrrrrrrl@{}}
\toprule
 & & & \multicolumn{2}{c}{Flow-level} & \multicolumn{2}{c}{Vuln-level}
 & \multicolumn{4}{c}{pass@$k$} \\
\cmidrule(lr){4-5}\cmidrule(lr){6-7}\cmidrule(lr){8-11}
Class & Flows & Att. & $n$ & SR & $n$ & SR & @1 & @2 & @3 & CI \\
\midrule
CWE-22  &  91 & 272 & 47 & 17.3 & 45 & 16.5 & 17.2 & 26.7 & 33.0 & [24.2, 43.1] \\
CWE-23  &  14 &  42 &  6 & 14.3 &  6 & 14.3 & 14.3 & 19.0 & 21.4 & [7.6, 47.6] \\
CWE-77  &  11 &  32 &  5 & 15.6 &  4 & 12.5 & 15.2 & 27.3 & 36.4 & [15.2, 64.6] \\
CWE-78  &  49 & 147 & 11 &  7.5 &  7 &  4.8 &  7.5 & 11.6 & 14.3 & [7.1, 26.7] \\
CWE-79  &   7 &  21 &  1 &  4.8 &  1 &  4.8 &  4.8 &  9.5 & 14.3 & [2.6, 51.3] \\
\midrule
Overall & 172 & 514 & 70 & 13.6 & 63 & 12.3 & 13.6 & 21.1 & 26.2 & [20.2, 33.2] \\
\midrule
Benign baseline & 172 & 516 & 0 & 0.0 & 0 & 0.0 & 0.0 & 0.0 & 0.0 & [0.0, 2.2] \\
\bottomrule
\end{tabular}
\end{table}

Our results show \textbf{most silencing rewrites also remove the
vulnerability-level finding} (the flow-level and vuln-level \textsc{SR}
columns in Table~\ref{tab:bsr}). Of the 70 confirmed silencing rewrites,
63 ($90.0\%$) leave no finding from the targeted rule through the
rewritten function. Accordingly, the overall vulnerability-level SR is
only $1.3$ percentage points below the flow-level SR ($12.3\%$ versus
$13.6\%$). A silencing rewrite therefore usually suppresses more than the
single path selected as its target. This relationship differs
substantially by class. For path traversal, 51 of 53 silencing rewrites
also silence the vulnerability level. The two rates are $17.3\%$ versus
$16.5\%$ for CWE-22 and both $14.3\%$ for CWE-23. In contrast, only 7 of
the 11 CWE-78 silencing rewrites suppress every same-rule finding through
the rewritten function, yielding vulnerability- and flow-level rates of
$4.8\%$ and $7.5\%$. In these command-injection cases, a tainted argument
typically reaches several shell sinks, so a single-function rewrite
removes only one.

Moreover, we observe that \textbf{silencing rewrites are cheap to
produce}. All 514 attempts together cost only about \$23 in model
inference, that is, \$0.13 per evaluated flow under the three-attempt
budget and about \$0.51 for each silenced flow. The same \$23 is the cost
of running the pipeline once as a robustness regression test over these
flows (Section~\ref{subsec:implications}). Because pass@$k$ had not
saturated by the third attempt, the cost per silenced flow would likely
fall further under a larger budget.

Finally, \textbf{the same generator silences nothing when it does not see
the alert} (the last row of Table~\ref{tab:bsr}). Under the same budget,
the benign baseline silences 0 of 516 valid rewrites and 0 of 172 flows,
with Wilson 95\% upper bounds of $0.7\%$ and $2.2\%$, against 70 of 514
and $26.2\%$ for the alert-aware generator on the same functions. The
benign rewrites are not no-ops. They change a median of 14 lines, with
early returns in place of nested conditionals, hoisted duplicated calls,
named intermediate values, and merged regular expressions, and CodeQL
reports the original flow on every one of them. The silencing rewrites
are in fact smaller, with a median of 3 changed lines. What separates the
two conditions is therefore not how much is changed but whether the
change lands on a modeling gap, and the 26.2\% measures what an edit
made with the alert in view can do rather than what refactoring does on
its own. One property of the baseline limits the comparison. The three
benign attempts on a function are usually the same refactoring, so
benign pass@$k$ does not grow with $k$, whereas the main run draws its
diversity from the alert-guided choice of location and construct.

\begin{keymsg}
\textbf{RQ1 Answer Summary}
With three natural, vulnerability-preserving rewrites per flow from a
bounded generator, CodeQL stops reporting $26.2\%$ of the flows it
originally detects. Silencing keeps rising through the third attempt with
diminishing gains. Path traversal is more exposed than command injection,
most silenced flows lose every finding of the targeted rule through the
rewritten function, and each silenced flow costs about \$0.51 to produce.
The same generator refactoring the same functions without the alert
silences none of them, so silencing depends on where an edit lands, not
on the fact of editing.
\end{keymsg}

\subsection{RQ2: Mechanisms Behind Silencing}
\label{subsec:rq2}

Every one of the 70 confirmed silencing rewrites traces to a single
construct at which CodeQL stops propagating taint. The controlled
experiments of Section~\ref{subsec:verify} establish the attribution. For
each case, restoring the blamed construct alone, with every other change of
the rewrite kept, makes CodeQL report the targeted flow again, and a
construct-level probe with the same source and sink shows the silenced
shape and its modeled counterpart side by side. Table~\ref{tab:gaps} lists
the seven gaps. Sixty-four of the 70 cases cross a gap inside an API that
CodeQL already models, three reach an API it does not model at all, and
three fall in the core analysis rather than in a library model. This
answers the question left open in Section~\ref{subsec:bg-taint}. A
pattern-based analyzer misses any unmodeled library call, so one might
expect silencing rewrites to route through obscure APIs. Instead, the
dominant weakness is narrow coverage gaps inside operations CodeQL already
models.

\begin{table}[t]
\centering
\caption{Modeling gaps behind the 70 confirmed silencing rewrites. For every
case, restoring the named construct alone, with the rest of the rewrite
kept, makes CodeQL report the targeted flow again. The 2.27.3 column counts
the cases that still silence CodeQL 2.27.3 when the same rewrites are re-run
under it (Section~\ref{subsec:impl}).}
\label{tab:gaps}
\small
\begin{tabularx}{\columnwidth}{@{}lXlrrl@{}}
\toprule
Gap & Construct at which taint is lost & Where & Cases & 2.27.3 & CWE \\
\midrule
F1 & \texttt{new URL(...)}: no taint on \texttt{.pathname} or
     \texttt{.href} (legacy \texttt{url.parse} is modeled)
   & modeled API & 43 & 43 & 22: 37, 23: 6 \\
F2 & \texttt{join} with a separator the query does not model
     (\texttt{"/"}, \texttt{path.sep}, \texttt{"|"}, \texttt{"\&\&"})
   & modeled API & 10 & 10 & 22: 3, 78: 7 \\
F3 & \texttt{join} on a receiver that is not an array literal
     (\texttt{concat}, \texttt{push}-built, helper parameter)
   & modeled API & 7 & 7 & 77: 5, 78: 2 \\
F4 & modeled built-in invoked through \texttt{.apply}, or its array
     argument copied element-wise before \texttt{join}
   & modeled API & 4 & 4 & 22: 2, 78: 2 \\
F5 & \texttt{path.format(\{dir, base\})} has no model, while
     \texttt{path.join} does
   & unmodeled API & 3 & 3 & 22: 3 \\
F6 & computed dispatch \texttt{table[key]()} to the sink
   & core analysis & 2 & 2 & 22: 2 \\
F7 & \texttt{reduce} accumulator written through a computed key
   & core analysis & 1 & 1 & 79: 1 \\
\midrule
\multicolumn{3}{@{}l}{Total} & 70 & 70 & \\
\bottomrule
\end{tabularx}
\end{table}

\textbf{The URL constructor (F1).}
CodeQL's URL model carries taint to a parsed URL's query components
(\texttt{.search}, \texttt{.searchParams}, \texttt{.hash}) but not to its
\texttt{.pathname} or \texttt{.href}. The omission is specific to the
WHATWG constructor \texttt{new URL}. The legacy \texttt{url.parse}
propagates taint to every component, and replacing \texttt{new URL} with
\texttt{url.parse} in a silenced rewrite, with nothing else changed, makes
the alert return. A rewrite that reads \texttt{new URL(input,
base).pathname} as a filesystem path therefore loses the taint at that
access. This is a true false negative, not a safe modeling choice.
Although \texttt{.pathname} looks like a benign component, the tainted
value survives into it and still reaches the sink, so the vulnerability is
preserved. This gap alone accounts for 43 of the 70
cases, all in path traversal.

\textbf{Array \texttt{join} (F2, F3).}
CodeQL recognizes \texttt{join} only under two conditions, and breaking
either drops the taint. First (F2), the separator must be the one each
query expects, \texttt{""} for path injection via
\texttt{StringConcatenation} and \texttt{" "} for command injection via
\texttt{UnsafeShellCommandConstruction}. Real code joins on other
separators just as naturally for the task, a path separator (\texttt{"/"}
or \texttt{path.sep}) for filesystem paths or a shell operator
(\texttt{"|"} or \texttt{"\&\&"}) for command pipelines, and none of these
is modeled. Second (F3), the array must be a literal whose elements the
model can enumerate, so producing it with \texttt{concat}, building it up
with \texttt{push}, or passing it through a function parameter drops the
taint. The two gaps account for 17 cases across path traversal and
command injection.

\textbf{Indirect invocation and element-wise copies (F4).}
Four cases keep a modeled call but change how its arguments arrive. Two
invoke \texttt{path.join} through \texttt{.apply}, which the summary for
\texttt{path.join} does not cover, whereas the equivalent direct call is
reported. Two copy a tainted argument array element by element, through
\texttt{push}, spread, or \texttt{concat}, before joining it. CodeQL
carries the taint on the array as a value through \texttt{join} but not
through the element-wise copy.

\textbf{An unmodeled sibling (F5).}
Three cases replace \texttt{path.join}, which is modeled, with
\texttt{path.format(\{dir, base\})}, which is not. This is the only gap in
a wholly unmodeled API, and the API is a sibling of a modeled function in
the same module.

\textbf{Core analysis (F6, F7).}
Three cases come from the core analysis rather than a library model. In
two, the call site is a dispatch table (\texttt{resolvers[kind]()}), so
which function runs is decided at runtime and CodeQL cannot follow the
taint into it. In one, the rewrite builds an options object with
\texttt{reduce}, assigning each field through a computed key
(\texttt{acc[key] = opts[key]}), then reads it back by name. CodeQL
follows a computed-key write on its own, and a \texttt{reduce} on its own,
but not the two together. These are genuine losses, though rare.

Seven of the 70 cases silence the targeted flow but leave another finding
of the same rule in the rewritten function (Section~\ref{subsec:rq1}).
Each still crosses one of the gaps above. What survives is a different
flow of the same rule, typically the same tainted argument reaching a
second sink, so for these seven the function stays flagged and only the
targeted flow is lost. They count as flow-level silencing only, which is
why the vulnerability-level SR of Section~\ref{subsec:rq1} is the
stricter measure of an alert that disappears outright.

\textbf{Which rewrites reach the gaps.}
We classify the 514 valid attempts by their transformation strategy.
Following the open-coding procedure of Section~\ref{subsec:audit}, the
authors inductively derived eight categories (Table~\ref{tab:strategy}).
For rewrites that combine several transformations, the dominant category
is the one that best explains the resulting dataflow change. Gaps and
strategies align closely. Every F1 case is parser routing (A), F2 and F3
are array assembly with \texttt{join} (D), F5 is helper indirection (B),
F6 is a dynamic call (C), and F7 is object indirection (E). The four F4
cases split between B, for the \texttt{.apply} calls, and D, for the
element-wise copies.

Effectiveness ranges widely across the eight strategies. The strategy with
the highest observed SR is URL / path-parser routing (A), which succeeds
on $41.0\%$ of its 105 attempts. Helper and library-call indirection (B)
shows a similarly high observed rate ($27.8\%$) on far fewer attempts. At
the other end, rewrites that merely reshape the data, through object or
map indirection (E, $1.4\%$), intermediate-variable hops (F, $0.0\%$), or
destructuring (G, $0.0\%$), almost never succeed, because CodeQL tracks
taint accurately through ordinary language constructs. We report these as
observed per-strategy rates rather than causal effects, since strategies
were not randomly assigned to comparable flows. Strategy A in particular
is confined to path-traversal cases, which Section~\ref{subsec:rq1}
already shows are more exposed. Array assembly with \texttt{join} (D) is
especially informative. It is the generator's most frequently selected
strategy, accounting for 176 attempts, but succeeds on only $10.8\%$.
Merely introducing a library call is therefore insufficient. Success
depends on whether the \emph{specific invocation and dataflow pattern}
fall within CodeQL's model. Because B, C, and H contain at most 18
attempts each, their SRs are indicative rather than precise comparisons
among strategies, though the cases we root-cause within them still point
to concrete mechanisms. The gaps also explain the per-class differences
of Section~\ref{subsec:rq1}. Path traversal can reach the URL gap and the
\texttt{join} gaps, whereas command injection reaches only
\texttt{join}.

\begin{table*}[htbp]
\centering
\caption{Rewrite strategies and their flow-level SR. \emph{n} counts
attempts and \emph{Sil.} confirmed silencing rewrites.}
\label{tab:strategy}
\small
\begin{tabularx}{\textwidth}{@{}llXrrr@{}}
\toprule
 & Strategy & Description & $n$ & Sil. & SR \\
\midrule
A & URL / path-parser routing
  & Read a parsed component off a URL or path parser.
  & 105 & 43 & 41.0 \\

B & Helper / library-call indirection
  & Wrap the call in a helper or invoke it indirectly.
  & 18 & 5 & 27.8 \\

C & Dispatch-table / dynamic call
  & Call the operation via a runtime-selected dispatch entry.
  & 12 & 2 & 16.7 \\

D & Array assembly + join
  & Assemble the string from array elements via \texttt{join}.
  & 176 & 19 & 10.8 \\

E & Object-property / map indirection
  & Store the value in an object or map property, read back by name.
  & 70 & 1 & 1.4 \\

F & Intermediate-variable hop
  & Relay the value through extra local variables.
  & 45 & 0 & 0.0 \\

G & Destructuring rebinding
  & Rebind the value via a destructuring pattern.
  & 74 & 0 & 0.0 \\

H & Other structural transform
  & Other structural refactorings not captured above.
  & 14 & 0 & 0.0 \\
\midrule
\multicolumn{3}{@{}l}{Overall}
  & 514 & 70 & 13.6 \\
\bottomrule
\end{tabularx}
\end{table*}

Moreover, we find that the rewrites are often small and predominantly
additive. The median rewrite changes 3 lines (mean 4.3). Across all
attempts, additions and in-place modifications account for 839 and
1{,}171 changed lines, compared with only 203 deletions. Rather than
removing the source or sink, the generator usually inserts one additional
transformation through which the tainted value passes. In silencing cases,
this inserted step is the URL parser, \texttt{join}, or indirect call that
crosses one of the gaps above. The resulting diff can therefore resemble
an ordinary local refactoring.

\begin{table}[t]
\centering
\caption{Source and sink locations: outside rewritten function
(OUT), inside but unchanged (KEEP), rewritten (REWR).}
\label{tab:endpoints}
\small
\begin{tabular}{@{}lrrrr@{}}
\toprule
 & \multicolumn{3}{c}{Sink} & \\
\cmidrule(lr){2-4}
Source & OUT & KEEP & REWR & Total \\
\midrule
OUT   & 30 & 26 & 37 & 93 \\
KEEP  & 4 & 34 & \textbf{167} & 205 \\
REWR  & 12 & \textbf{172} & 32 & 216 \\
\midrule
Total & 46 & 232 & 236 & 514 \\
\bottomrule
\end{tabular}
\end{table}

Table~\ref{tab:endpoints} records whether the source and sink lines fall
outside the rewritten function (OUT), inside but unchanged (KEEP), or
rewritten (REWR). Of the 405 attempts with both endpoints inside the
function, 339 ($83.7\%$) rewrite exactly one of them, and the 32 that
rewrite both never succeed. Which endpoint is the effective one tracks the
class. Path traversal is silenced by rewriting the source, routing input
through a URL or path API into the \texttt{.pathname} gap, and command
injection by rewriting the sink, assembling the command via
\texttt{join}. A silencing rewrite is thus a small change at the one
endpoint that exposes its class's modeling gap.

\textbf{Persistence across releases.}
The gaps are not confined to the specific release we study. Re-running the
pipeline under CodeQL 2.27.3 (Section~\ref{subsec:impl}) changes no
outcome. On the original packages the newer release still reports 172 of
the 173 targets, the exception being the minified flow already excluded
from the corpus. On the rewrites, all 70 confirmed silencing rewrites still
silence it, with the per-gap split of Table~\ref{tab:gaps}, and none of
the other 444 rewrites loses its alert. The construct-level probes that
establish the attribution at the start of this section behave the same
way. Their 73 one-line variants receive the same alert-or-silent verdict
under both releases. In 2.27.3 the \texttt{TaintedPath} query still
carries no taint from \texttt{new URL(u, base)} to \texttt{.pathname} or
\texttt{.href}, while the legacy \texttt{url.parse} components remain
modeled. Every gap in
Table~\ref{tab:gaps} therefore persists unchanged from 2.22.4 to 2.27.3.

\begin{keymsg}
\textbf{RQ2 Answer Summary}
Every silencing rewrite traces to one construct at which CodeQL stops
propagating taint, and restoring that construct alone brings the alert
back. Of the 70 cases, 64 cross a gap inside an API CodeQL already
models, 3 reach an unmodeled sibling of a modeled function, and 3 fall in
the core analysis. The single largest gap is the WHATWG URL constructor,
whose \texttt{.pathname} carries no taint although the legacy
\texttt{url.parse} is modeled, and it alone accounts for 43 of the 70. The
rewrites that reach these gaps are small, additive, and usually change
only one endpoint, so they resemble routine refactorings. All seven gaps
persist unchanged in CodeQL 2.27.3.
\end{keymsg}

\subsection{RQ3: Generality Beyond CodeQL}
\label{subsec:rq3}

For each transfer analyzer we evaluate only the advisories VulcaN labels
Score-A for it (Section~\ref{subsec:dataset}) and that the analyzer still
detects on the original package, since a rewrite can silence an analyzer
only where it originally detected the vulnerability. This leaves ODGen
with 93 of the 172 flows and njsscan with 63, spanning 278 and 188 rewrite
attempts (one attempt in each scope is invalid). Both scopes are dominated
by path traversal, with command injection and XSS too thin to read on
their own (Section~\ref{subsec:ttv}), so the findings below are
effectively about path traversal. The two sides use different
granularities. The CodeQL side reuses the Stage~3 confirmed silencing
rewrites, whereas a transfer analyzer counts as detecting the vulnerability
whenever it reports the mapped class anywhere in the rewritten package, a
generous rule that makes the reported transfer rates conservative. A
silencing rewrite \emph{transfers} to an analyzer that flagged the
vulnerability on the original code but not on the rewrite. As a control,
we measure each analyzer's miss rate on the rewrites that did not silence
CodeQL, which gives a baseline for its own sensitivity to rewriting.

Within the evaluated scopes, the rewrites that silence CodeQL expose a
modeling boundary that is not specific to CodeQL. Most of them also
silence ODGen, a separately designed semantic taint analyzer, even though
they were never targeted at it, whereas the pattern-based scanner njsscan
behaves differently. Table~\ref{tab:joint} splits each analyzer's
detection by CodeQL's verdict. Of the 43 silencing rewrites in ODGen's
scope, 32 ($74\%$, 95\% CI $[60, 85]$) also silence it, against only
$24\%$ ($57/235$, CI $[19, 30]$) of the controls, a gap of 50 percentage
points.

\textbf{ODGen.}
The transfer is concentrated in a single strategy. Table~\ref{tab:xstrategy}
groups the rewrites by the strategy taxonomy of Section~\ref{subsec:rq2},
and each cell is the rewrites an analyzer missed over the total in that
strategy and column. Parser routing (strategy~A) supplies 31 of ODGen's 32
transfers. ODGen misses $86\%$ ($31/36$) of these silencing rewrites
against only $2\%$ ($1/53$) of the matching controls. The rewrites read the
path from the WHATWG constructor, \texttt{new URL(req.url, base).pathname},
which CodeQL does not model as taint-carrying in any form. ODGen splits
sharply by syntax instead. It catches the 5 rewrites that read
\texttt{.pathname} inline off the constructor and misses the other 31, 29
of which bind the URL to a variable and read the property in a later
statement. ODGen therefore does model the getter. What it loses is the
taint on a property read off a stored object. The bulk of ODGen's misses
on both sides trace to this one step, including the destructuring
controls (strategy~G) discussed below, with only a scattered handful
elsewhere (strategy~D). Outside these three strategies, ODGen misses none
of the remaining 59 controls, so its misses concentrate on this one step
rather than spreading across strategies.

\begin{table}[t]
\centering
\caption{Misses per strategy on the rewrites, by transfer analyzer:
silencing rewrites (\emph{transfer}) versus rewrites that did not silence
CodeQL (\emph{control}). Each cell is misses over rewrites in scope.}
\label{tab:xstrategy}
\small
\begin{tabular}{@{}lcccc@{}}
\toprule
 & \multicolumn{2}{c}{ODGen} & \multicolumn{2}{c}{njsscan} \\
\cmidrule(lr){2-3}\cmidrule(lr){4-5}
Strategy & transfer & control & transfer & control \\
\midrule
A\, parser routing & 31/36 & 1/53   & 17/29 & 29/53 \\
D\, array+join     & 1/6   & 7/70   & 2/3   & 8/33  \\
G\, destructuring  & ---   & 49/53  & ---   & 31/35 \\
Other              & 0/1   & 0/59   & 0/1   & 7/34  \\
\midrule
All                & 32/43 & 57/235 & 19/33 & 75/155 \\
\bottomrule
\end{tabular}
\end{table}

\begin{table}[t]
\centering
\caption{Detection on the rewrites, by CodeQL (rows) and the second tool
(columns). \emph{det}: detected, \emph{miss}: not detected.}
\label{tab:joint}
\small
\begin{subtable}{0.49\columnwidth}\centering
  \caption{ODGen}
  \begin{tabular}{@{}lrr@{}}
  \toprule
   & det & miss \\
  \midrule
  CodeQL miss & 11 & \textbf{32} \\
  CodeQL det & 178 & 57 \\
  \bottomrule
  \end{tabular}
\end{subtable}\hfill
\begin{subtable}{0.49\columnwidth}\centering
  \caption{njsscan}
  \begin{tabular}{@{}lrr@{}}
  \toprule
   & det & miss \\
  \midrule
  CodeQL miss & 14 & \textbf{19} \\
  CodeQL det & 80 & 75 \\
  \bottomrule
  \end{tabular}
\end{subtable}
\end{table}

\textbf{njsscan.}
Because njsscan matches a syntactic pattern around the sink rather than
following a flow, a flow-modeling boundary gives it no special difficulty.
It misses the silencing rewrites ($58\%$, $19/33$, 95\% CI $[41, 73]$)
only slightly more often than the controls ($48\%$, $75/155$, CI
$[41, 56]$), a gap far smaller than ODGen's, and the two intervals
overlap widely, so we find little evidence that a silencing rewrite is
harder for njsscan than a control. Its misses therefore track
rewriting in general rather than the silencing of CodeQL specifically.
This locates the transfer in flow-based semantic analysis and confirms
that our transfer measure does not simply credit every miss as transfer.

\textbf{A shared boundary.}
The two flow-based analyzers lose strategy~A at the same stored-object
read for unrelated internal reasons, one missing the parser model and the
other the read off a bound object. Strategy~G confirms these are distinct
mechanisms rather than one shared bug, since it virtually never silences
CodeQL yet still silences ODGen at that same read, which misses $92\%$
($49/53$) of its controls. What recurs across the two designs is the
operation, routing a tainted value through a library parse and reading a
component back off a stored object, not any single model defect.

\textbf{Stacking analyzers.}
Running more analyzers does not recover the detections CodeQL loses on
most of these rewrites. This follows from the boundary being shared rather
than tool-specific. Because ODGen loses strategy~A at the same
stored-object read, adding it to CodeQL catches only the inline cases both
already handle. Even CodeQL, ODGen, and njsscan together miss $53\%$
($16/30$, 95\% CI $[36, 70]$) of the silencing rewrites that lie in both analyzers' common
scope, with njsscan recovering 9, ODGen 2, and both 3. Since an analyzer
counts as catching one whenever it flags the vulnerability anywhere, this
understates how many rewrites truly escape. The two analyzers catch
different rewrites, and njsscan's larger share is not a stronger signal,
since it detects the controls as often. Even that partial recovery is
fragile, resting on the same inline-only handling that the bound-variable
rewrites already defeat. Closing the gap calls for changing the analysis
rather than adding detectors, by modeling the constructs they miss or by
canonicalizing such rewrites into a form those models cover.

\begin{keymsg}
\textbf{RQ3 Answer Summary} Within the evaluated, path-traversal-dominated
scopes, the rewrites that silence CodeQL expose a modeling boundary shared
across analysis designs rather than a CodeQL-specific limitation. Most of
them ($74\%$) also silence the independently built semantic analyzer
ODGen, mostly at the same stored-object read, whereas the pattern-based
njsscan misses silencing rewrites and controls alike, which locates the
transfer in flow-based analysis. Because the boundary is shared, all three
detectors together still miss $53\%$ of the common-scope silencing
rewrites, so closing it requires better modeling or normalization rather
than more tools.
\end{keymsg}

\section{Discussion}
\label{sec:discussion}

\subsection{Implications}
\label{subsec:implications}

\textbf{For SAST users, a disappearing alert after an ordinary-looking
maintenance change is not reliable evidence that the vulnerability has
been fixed, even when multiple analyzers agree.}
Prior work has shown that practical static analyses are necessarily
``soundy'' and that users may underestimate false negatives and
undocumented tool limitations
\cite{bonett2018discovering,ami2024false}.
Our results expose a more operational failure mode: a tool may correctly
report a real source-to-sink flow, yet a natural maintenance rewrite can
move the same vulnerable behavior across a modeling boundary and make
the alert disappear. Edits made without the alert in view rarely have
this effect, as the benign baseline of RQ1 silences none of the same
flows. The risk arises when the edit is made with the alert in view, as
in alert-driven fixing. A clean re-scan should therefore be accompanied
by
independent evidence of vulnerability removal, such as a security
regression or exploit test, or manual source-to-sink inspection,
especially when the change touches the reported flow. The gate and the diff
it admits are the same for all three producers of Section~\ref{sec:threat},
so this advice covers a contributor who keeps the vulnerability deliberately
as much as an assistant or developer who does so unknowingly.

\textbf{For SAST developers, the most immediate improvement target is
partial coverage within library APIs already modeled.}
It is well established that taint analyzers depend on library specifications, and prior work has proposed techniques for
inferring or extending such specifications for JavaScript libraries
\cite{staicu2020extracting,inspectjs}.
Our RQ2 root-cause analysis identifies a narrower and more actionable
problem: gaps within otherwise modeled APIs. By mapping each silenced flow
to the modeling gap it crosses, we provide a concrete set of repair
targets. Some are local, such as propagating taint to a parsed URL's
\texttt{pathname}, although such an omission may be a deliberate precision
choice that a fix would have to revisit. Others require more general models,
since patching individual syntax patterns leaves equivalent rewrites
uncovered. Our pipeline can also be reused as a robustness regression test
for library models, run whenever models or queries change.

\textbf{For the research community, robustness to natural,
vulnerability-preserving rewrites should complement conventional SAST
metrics such as detection rate, precision, recall, and benchmark
coverage.}
RQ1 shows that such variants are cheap to generate and require no
knowledge of the analyzer's internals, so robustness testing can be a
routine part of evaluation rather than a one-time audit. Moreover, RQ3
shows that the exposed boundary is not specific to one tool: the same
rewrites silence an independently designed analyzer at the same operation,
but for different internal reasons. Future SAST
evaluations should therefore measure robustness across natural,
behavior-preserving variants, rather than testing only one canonical
representation of each vulnerability.

\subsection{Threats to Validity}
\label{subsec:ttv}
\textbf{Construct validity.}
Our three constructs, detection loss, vulnerability preservation, and
naturalness, are defined in Sections~\ref{sec:threat}
and~\ref{subsec:verify}. We judge preservation at the level of taint flow,
as the dataset's annotations do, and did not construct exploits, so a
preserved flow is one the analyzer should report rather than one shown to
be exploitable. Naturalness is the most subjective construct. We judge it
against the two criteria of Section~\ref{subsec:audit}, and the judges are
the authors, who also see each attempt's outcome. We did not have
maintainers or blinded raters review the rewrites, so our results show
that the rewrites satisfy the stated constraints, not how real reviewers
would act on them. The benign baseline uses the same LLM, so it measures
refactoring without knowledge of the alert rather than human maintenance
commits.

\textbf{Internal validity.}
A missing CodeQL finding could stem from invalid code, an incomplete run, a
flow-matching error, or an unintended semantic change rather than a genuine
blind spot. The verification procedure of Section \ref{subsec:verify} guards against
these, but the flow alignment and root-cause steps remain manual, so residual
human error cannot be excluded.

\textbf{Conclusion validity.}
Because the verification cascade re-audits every silencing rewrite but only
samples the rest, it can miss real silencing rewrites but rarely labels a
spurious one, so the reported rate is a floor, not a ceiling. A few
conclusions also rest on limited statistical support, as CWE-77 and CWE-79 contain few flows and the
cross-tool transfer is estimated from 30 shared silencing rewrites, so we
report these subgroup and transfer rates with confidence intervals and read
them as indicative rather than conclusive.

\textbf{External validity.}
Our subjects are npm/JavaScript and the gaps lie in CodeQL's JavaScript models, so our findings may not carry over to other ecosystems, languages, or vulnerability classes. The corpus is dominated by path traversal and command injection, and
because the other analyzers detect little beyond path traversal the
cross-tool findings of Section~\ref{subsec:rq3} are effectively
path-traversal-only. This imbalance traces to the scarcity of
\texttt{(source, sink)}-annotated data (Section~\ref{subsec:dataset}), a
prerequisite for broadening this work. The measured rate also reflects one
bounded rewrite generator (Section~\ref{sec:threat}), so a stronger model
or a larger budget would likely raise it, the rate is best read as a lower
bound, and the mix of strategies in Section~\ref{subsec:rq2} may differ
for other models. Finally, the gaps sit in CodeQL v2.22.4. Re-running every
rewrite under CodeQL 2.27.3 changes no outcome (Section~\ref{subsec:rq2}),
and the same constructs silence a differently built analyzer, so what we
surface is a modeling boundary shared across releases and implementations
rather than a version-specific bug.

\section{Related Work}
\label{sec:related}

\textbf{Semantics-preserving transformations against code analysis.}
Semantics-preserving transformations have long defeated automated detectors, from classical malware obfuscation~\cite{moser2007limits} to adversarial examples for learned code models~\cite{yefet2020adversarial,zhang2020generating}, later optimized for naturalness~\cite{yang2022natural} or aimed at code-representation and authorship models~\cite{quiring2019misleading}. Recent work uses LLMs to generate
such code directly: CodeBreaker disguises injected vulnerabilities in
code-completion backdoors to evade SAST and LLM-based
detectors~\cite{yan2024llm}, EaTVul targets deep-learning vulnerability
detectors~\cite{liu2024eatvul}, and, most closely related, Pagano et al.\
iteratively obfuscate injected JavaScript until static analyzers stop reporting
the vulnerability~\cite{pagano2026obfuscating}. We instead start from confirmed
flows CodeQL already detects, require each rewrite to preserve their source,
sink, interface, and vulnerable behavior as a plausible maintenance change,
and locate the modeling gap behind each silenced flow and its zero-shot
transfer to independently designed analyzers.

\textbf{Weaknesses of static taint analysis.}
Production taint analyzers deliberately trade soundness for precision,
scalability, and usability, a tradeoff known as
soundiness~\cite{livshits2015soundiness}. The resulting gaps can be large.
On real npm packages, even the three best-performing SAST tools combined
detect only 57.6\% of VulcaN's annotated vulnerabilities~\cite{vulcan}, and
static C analyzers miss between 47\% and 80\% of known vulnerabilities in
real programs~\cite{lipp2022empirical}. A recurring source of missed flows is
library and API modeling, where a missing or incomplete summary of how taint
propagates among arguments, receivers, and results can terminate an otherwise
valid flow. Prior work infers such taint specifications automatically, from
Merlin's probabilistic inference~\cite{merlin} and large-scale
learning~\cite{seldon} to JavaScript-specific
techniques~\cite{inspectjs,bimodal} and LLM-driven extraction for npm
packages~\cite{semtaint}. These approaches treat the boundary as a coverage
problem: APIs without a model are the target, and the output is a new model.
We ask the complementary question of where the existing models give way under
ordinary code changes, and find that the answer lies mostly inside APIs that
already have a summary rather than in APIs that have none.

\textbf{Testing static analyzers.}
Metamorphic and differential testing find analyzer defects by checking that
a program and a transformed variant receive consistent
verdicts~\cite{segura2016survey}. The variants come from random program
generation~\cite{cuoq2012testing}, from differential comparison of several
analyzers on the same inputs~\cite{klinger2019differentially,taneja2020testing},
from semantics-preserving transformations of the analyzers' own test
programs~\cite{zhang2023statfier}, and, for security-focused analyzers, from
systematic mutation of synthetic Android apps that plants flows the analyzer
should report~\cite{bonett2018discovering}. These studies target soundness
and precision in general, drive the transformations with fixed rules or
synthetic programs, and report defects as a count. We apply the same
relation to real, confirmed vulnerabilities, generate the variants with an
LLM under a naturalness requirement so that each violation is a change a
maintainer could plausibly make, and attribute every violation to the part
of the modeling boundary it crosses.

\section{Conclusion}
\label{sec:conclusion}
We tested whether CodeQL's taint tracking stays robust when vulnerable code
is rewritten rather than fixed, using an LLM that sees only the alert
CodeQL exposes and produces natural, vulnerability-preserving rewrites of
one function. Three such rewrites per flow silence 26.2\% of the flows
CodeQL originally detects on real npm vulnerabilities, at about \$0.51 per
silenced flow and with the vulnerability verified to survive in every
case. Every silencing rewrite traces to one of seven modeling gaps, and
restoring the responsible construct alone brings the alert back. The
decisive weakness is not the APIs CodeQL leaves unmodeled but narrow gaps
inside the ones it does model, led by a URL constructor whose
\texttt{pathname} carries no taint. The same boundary silences an
independently built semantic analyzer, so closing it requires changing the
models rather than stacking detectors. For users, a disappearing alert
after an ordinary-looking change made with the alert in view is weak
evidence that the flaw is gone, while the same generator refactoring
without the alert silences none of the flows.
For analyzer developers, the gaps are concrete repair targets, and the
pipeline can serve as a regression test for library models. As
LLM-assisted maintenance makes such rewrites routine, robustness to them
belongs alongside detection rate in how taint analyzers are evaluated.

\noindent\textbf{Replication Package.} Our entire study artifact, including the evaluated flows, rewrites, and all tool outputs, is available at \url{https://figshare.com/s/63962213d3744750c45d}.

\bibliographystyle{ACM-Reference-Format}
\bibliography{refs}

\end{document}